\documentclass[10pt,conference]{IEEEtran}
\usepackage{url}
\usepackage[colorlinks=false, hidelinks]{hyperref}
\usepackage{framed}
\usepackage{fontawesome}
\usepackage{amsmath}
\usepackage{amssymb}
\usepackage{graphicx}
\usepackage{subcaption}
\usepackage{soul}
\usepackage{lipsum}
\usepackage{flushend}
\usepackage{booktabs}

\IEEEoverridecommandlockouts

\usepackage{xspace}
\newcommand{\ie}{\textit{i.e.},\xspace}
\newcommand{\eg}{\textit{e.g.},\xspace}

\newcommand{\etal}{et~al.\xspace}

\usepackage{xcolor}

\definecolor{formalshadelight}{RGB}{242,242,242}
\definecolor{formalshadedark}{RGB}{50,50,50}
\newenvironment{highlight}{%
  \MakeFramed{\advance\hsize-\width\FrameRestore}%
  \noindent\begin{minipage}{\linewidth}\noindent\hspace{-4.55pt}% disable indenting first paragraph
}
{%
  \vspace{2pt}\vspace{-2pt}\end{minipage}\endMakeFramed%
}

\ifCLASSINFOpdf

\else

\fi

\begin{document}

%\title{Green AI: A Comparative Study of Programming Languages}
\title{Green AI: Which Programming Language Consumes the Most?}

\author{\IEEEauthorblockN{Niccol\`o Marini\IEEEauthorrefmark{1}\thanks{\IEEEauthorrefmark{1} The first three authors contributed equally to this work.},
Leonardo Pampaloni\IEEEauthorrefmark{1},
Filippo Di Martino\IEEEauthorrefmark{1}, 
Roberto Verdecchia\IEEEauthorrefmark{2}\IEEEauthorrefmark{3}\thanks{\IEEEauthorrefmark{3} Corresponding author.},
and Enrico Vicario\IEEEauthorrefmark{2}}
\IEEEauthorblockA{University of Florence, Software Technologies Laboratory\\ \IEEEauthorrefmark{1}name.surname@edu.unifi.it, \IEEEauthorrefmark{2}name.surname@unifi.it}
%\IEEEauthorblockA{\IEEEauthorrefmark{2}University of Florence, Email: name.surname@unifi.it}
}


\maketitle

\begin{abstract}
AI is demanding an evergrowing portion of environmental resources. Despite their potential impact on AI environmental sustainability, the role that programming languages play in AI (in)efficiency is to date still unknown. With this study, we aim to understand the impact that programming languages can have on AI environmental sustainability. To achieve our goal, we conduct a controlled empirical experiment by considering five programming languages (C++, Java, Python, MATLAB, and R), seven AI algorithms (KNN, SVC, AdaBoost, decision tree, logistic regression, naive bayses, and random forest), three popular datasets, and the training and inference phases. The collected results show that programming languages have a considerable impact on AI environmental sustainability. Compiled and semi-compiled languages (C++, Java) consistently consume less than interpreted languages (Python, MATLAB, R), which require up to 54x more energy. Some languages are cumulatively more efficient in training, while others in inference. Which programming language consumes the most highly depends on the algorithm considered. Ultimately, algorithm implementation might be the most determining factor in Green AI, regardless of the language used. As conclusion, while making AI more environmentally sustainable is paramount, a trade-off between energy efficiency and implementation ease should always be considered. Green AI can be achieved without the need of completely disrupting the development practices and technologies currently in place.
\end{abstract}

% \fi
\IEEEpeerreviewmaketitle

\section{Introduction}
\label{sec:intro}
As artificial intelligence (AI) is becoming more and more ubiquitous, the resources required to train and deploy AI models have surged in recent times~\cite{Strubell2019,Lacoste2019}. While the development of sophisticated AI algorithms has led to remarkable innovations, it has also introduced new challenges related to their energy consumption and carbon emissions. In a world of limited environmental resources, the ever-growing energy required to power AI is now a pressing concern~\cite{Bender2021}.

The field of \textit{Green AI}~\cite{Schwartz2020} tackles this issue by studying how AI-centric applications can be designed, deployed, and used to make AI more environmentally sustainable. Among a rapidly growing corpus of literature, research focused primarily on topics such as AI footprint monitoring, tradeoffs between accuracy and energy consumption, energy efficient algorithm design, and sustainable AI model deployment~\cite{verdecchia2023systematic}.

By considering at large the software sustainability research landscape, some studies investigated the impact that programming languages have on energy consumption~\cite{pereira2021ranking, georgiou2017analyzing, pereira2017energy}. Results show that language choice plays a crucial role in the energy efficiency of software-intensive systems. 

Albeit both Green AI and environmental sustainability of programming languages are widely investigated subjects, the intersection of the two topics appears to date to be still an uncharted territory. In this research, we aim to make a first step into the domain opened by the question:

\begin{quote}
    ``\textit{What is the impact of programming languages on AI energy efficiency?}''
\end{quote}

To answer the question, we present an empirical experiment considering popular AI libraries of five programming languages (Python, C++, Java, R, and MATLAB), seven exemplary AI algorithms (k-nearest neighbors, support vector machine, AdaBoost, decision tree, logistic regression, naive bayes, and random forest), and three broadly used AI datasets (Iris, Breast Cancer, and Wine Quality). 

Results show that choosing one language over another can have up to a 54x increase in energy consumption. Additionally, programming language energy efficiency differs when considering the training and inference phases, making their adoption to improve software sustainability highly depend on the application context considered. Finally, as a conjecture based on the empirical measurements collected, ultimately the most important factor influencing AI energy efficiency could be how algorithms are implemented, rather than the specific algorithm or programming language considered. This consideration may hold true even if, despite potentially using pre-compiled libraries, the overhead of interpreted languages cannot be fully offset by code optimization.

The main contributions of this study are:
\begin{itemize}
    \item A rigorous empirical comparison of AI energy consumption across multiple programming languages, 
    \item A statistical and discursive interpretation of the results collected \textit{via} the empirical experiment,
    \item A comprehensive replication package, including all data, scripts, and settings necessary to reproduce in their entirety the presented results.\footnote{Replication package: \url{https://doi.org/10.6084/m9.figshare.27645786}}
\end{itemize}

\section{Experimental Methodology}
In this section we report all details regarding our experiment, including research goal and questions~(Section~\ref{sec:rqs}), experimental objects~(Section~\ref{sec:datasets}), dependent and independent experimental variables~(Section~\ref{sec:variables}), and experimental procedure~(Section~\ref{sec:procedure}).

\subsection{Research Goal and Questions}
\label{sec:rqs}
By considering the goal-question metric experiment definition by Basili~\etal~\cite{caldiera1994goal}, our goal is formulated as follows: 

\noindent
    \textit{Analyze} AI model training and inference\\
    \textit{For the purpose of} understanding the energy consumption\\
    \textit{With respect to} programming languages\\
    \textit{From the viewpoint of} software researchers and practitioners\\
    \textit{In the context of} artificial intelligence.

Our goal takes a pragmatic stance, by studying in a black-box fashion the energy consumed by popular AI libraries implemented in different programming languages that developers widely use ``in the wild''.

The research goal is directly translated into our main research question (RQ) guiding this study, namely:
\begin{LaTeXdescription}
    \item [$RQ_1$] \textit{What is the impact of programming languages on AI energy efficiency?}
\end{LaTeXdescription}

The research question is further decomposed into two sub-RQs to isolate the AI phase considered. Accordingly:
\begin{LaTeXdescription}
    \item [$RQ_{1.1}$] \textit{What is the impact of programming languages on AI training energy efficiency?}
\end{LaTeXdescription}

\begin{LaTeXdescription}
    \item [$RQ_{1.2}$] \textit{What is the impact of programming languages on AI inference energy efficiency?}
\end{LaTeXdescription}

With the two research questions, we aim to gain a comprehensive understanding on the impact that programming languages can have not only on the creation of AI models ($RQ_{1.1}$), but also on their follow-up utilization ($RQ_{1.2}$).

\subsection{Experimental Objects}
\label{sec:datasets}
For our experiment, we consider tree different datasets taken from the popular University of California Irvine Machine Learning Repository.\footnote{\url{https://archive.ics.uci.edu/ml/datasets}. Accessed 23rd October 2024.} The three datasets, used in the experiment for classification tasks, are (i) ``Iris'', comprising 150 tabular instances and four features, (ii) ``Breast Cancer Wisconsin (Original)'', comprising 699 multivariate instances and 30 features, and (iii) ``Wine Quality'', comprising 4.9K multivariate instances and~four~features.

The datasets are chosen by ensuring heterogeneity in terms of subject areas, number of instances, features, and data types. As further detailed in Section~\ref{sec:ttv}, the selection of the datasets is guided by a tradeoff between external and internal validity to ensure the rigor of the collected experimental data.

\subsection{Experimental Variables}
\label{sec:variables}
Our experiment is characterized by (i) a set of independent variables, namely programming languages, algorithms, and AI phase, the latter of which is also used as blocking factor, and (ii) one dependent variable, namely energy consumption. 

\subsubsection{Programming Languages} 
The focal independent variable of this study are programming languages, picked by considering their popularity and  AI applicability. The chosen languages are: \textit{Python}, \textit{C++}, \textit{Java}, \textit{R}, and \textit{MATLAB}.

\subsubsection{Algorithms}
To gain a comprehensive understanding on how programming languages can impact AI energy consumption, we consider a heterogeneous range of popular AI algorithms. The seven algorithms selected are \textit{k-nearest neighbors} (KNN), \textit{support vector classifier}~(SVC), \textit{AdaBoost}, \textit{decision tree}, \textit{logistic regression}, \textit{naive bayes}, and \textit{random forest}. 

Algorithms selection is guided by algorithmic heterogeneity, representativeness of different algorithmic families, popularity, and cross-language off-the-shelf availability in \textit{de facto} standard libraries such as \textit{scikit-learn}.\footnote{\url{https://scikit-learn.org}. Accessed 23rd October 2024.}

We purposely opt not to focus in this study on GPU-intensive algorithms such as deep-neural networks, due to their different algorithmic and hardware setting nature, leaving their consideration as future work (see Section~\ref{sec:conclusion}).

\subsubsection{AI phase (training and inference)} 
The blocking factor of our experiment are the AI training and inference phases. Experimental results are collected and analyzed separately for each of the two phases in order to answer our RQs (see Section~\ref{sec:rqs}). An 80-20 split is used to measure the energy consumption of the two phases.

\subsubsection{Energy Consumption}
Trivially, our dependent variable is the total energy, in Joules, consumed by all hardware components used to execute the computations set up in our experiment. Specifically, in the following we simply refer to \textit{energy consumption} as the sum of the energy required by all hardware components involved, namely CPU, GPU, and RAM. For details regarding the process used to collect energy measurements, refer to Section~\ref{sec:procedure}. 

We purposely refrain from considering as dependent variables other software sustainability metrics, \eg carbon footprint, as these metrics are highly influenced by external \textit{in vivo} factors, such as execution date, location, and energy availability, making results hard if not even impossible to be compared and reproduced.

% We avoid to consider other AI lifecycle phases, such as data curation and model fine-tuning, as highly prone to subjective biases YADAYADAYADA

\subsection{Experimental Setup}
\label{sec:procedure}
In this section we document the experimental setup, in terms of experimental hardware and software infrastructure utilized. 

The entirety of the experiment is executed on a machine featuring an Apple M2 chip with an 8-core CPU, an integrated 10-core GPU, 8GB of unified memory, 256GB SSD, running on macOS~v14.0. 

To collect energy measurements, we utilize the CodeCarbon library\footnote{\url{https://github.com/mlco2/codecarbon},. Accessed 23rd October 2024.} (v2.2.2), a Python library collecting unified memory architecture energy usage data based on a native utility package of the operating system used for the experiment.

Regarding algorithm implementation, to avoid potential threats to internal validity, we rely on popular open source off-the-shelf AI libraries available online. Specifically, for Python we use \textit{scikit-learn}~(v1.5.1), providing the source code of all algorithms considered. For C++ we use the implementation of all algorithms provided in the \textit{mlpack} library~(v3.8.6), except AdaBoost, which is found in the \textit{XGBoost} library~(v2.1.2). All Java algorithms are taken from the \textit{WEKA} library~(v4.5.0), with the exception of Naive Bayes, that is implemented in the \textit{OpenCV} library (v4.10.0). The implementation of MATLAB algorithms are taken from the \textit{Statistics and Machine Learning Toolbox} (v24.1). Finally, R algorithms are taken from the \textit{caret} library~(v6.0.94). Further software environment specifics, such as all dependencies of the packages reported above accompanied by their version number used in our experiment, are provided for replication and scrutiny purposes in our replication package~(see Section~\ref{sec:rep_pkg})

All algorithms are compiled and run in their default settings, by ensuring that all variables characteristic of the algorithms studied (\eg number of nearest neighbors for KNN), are set as equal across the different implementations considered.

\subsection{Experimental data collection procedure}
\label{sec:data_collection}
To mitigate potential threats to internal validity each experimental run, \ie combination of independent variables, is executed 30 times, leading to a total of 6.3k experimental executions conducted for this research. The median energy consumption collected across the 30 run repetitions is then utilized as representative sample of the experimental run. \textit{Via} this methodology, we ensured that our study design is statistically resilient to random confounding factors that may affect our measurements at runtime.

\begin{figure}[hbpt]
    \centering
    \includegraphics[width=1\columnwidth]{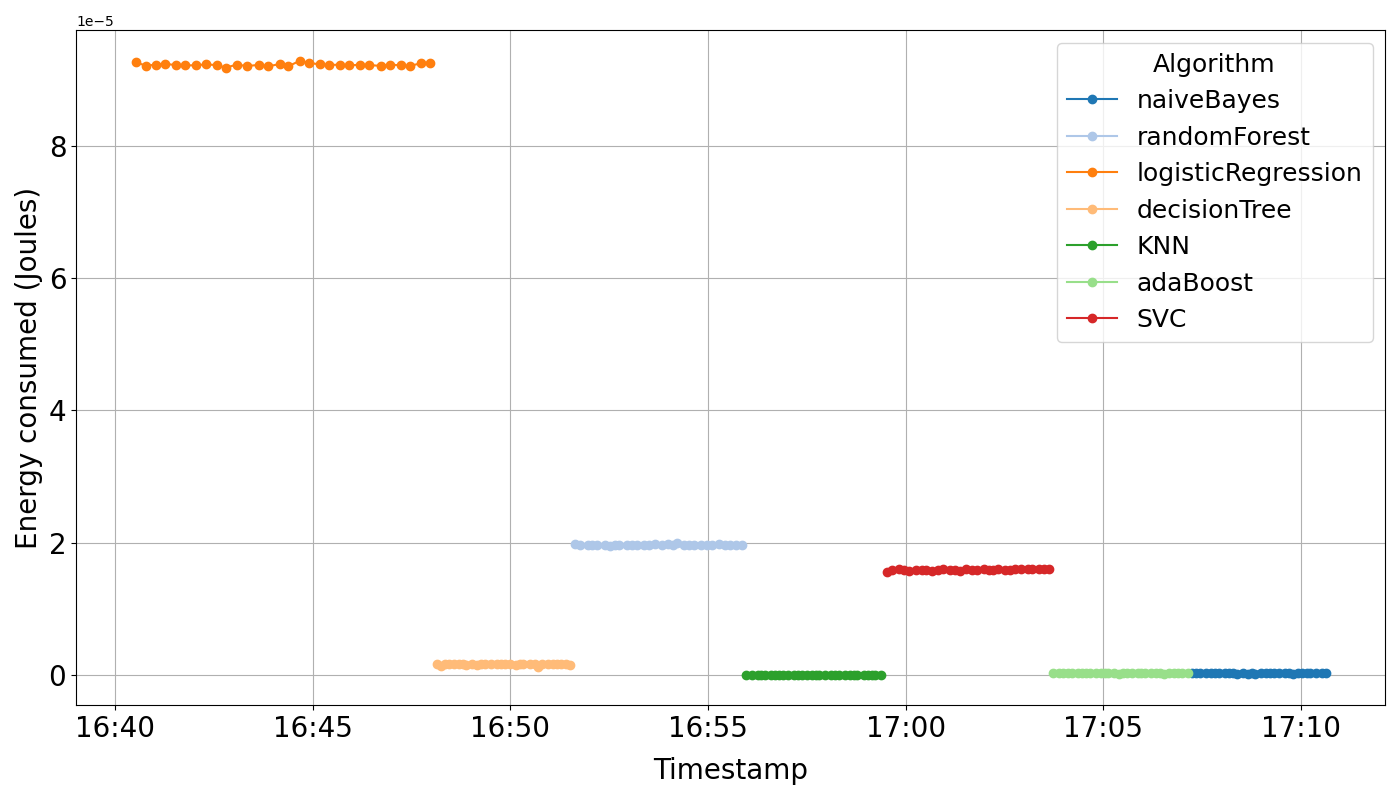}
    \caption{Example of energy time series used for results quality assurance}
    \label{fig:timeseries}
\end{figure}

To further mitigate the the influence of possible confounding variables affecting our experimental measurements, \eg unnoticed execution of a background process or CPU throttling, time series of all run repetitions are collected and analyzed to identify potential dependencies between executions. 
For completeness, an example of time series used for such quality assurance process is reported in Figure~\ref{fig:timeseries}.

The experiment is run in a controlled \textit{in vitro} environment, by enforcing that solely computations related to the experiment are run during the experimental measurements. 

Measurements are conducted by isolating at line of code level the training and inference operations of the utilized library. Specifically, dataset operations such as loading and splitting, and ancillary operations required to execute the train and inference operations, \eg code compilation and Java Virtual Machine startup time, are rigorously omitted to focus the energy measurements exclusively on algorithm executions. 

In order to gain comprehensive insights into the programming language energy consumption, the measurements collected across the three considered datasets are jointly analyzed for interpretation in the Results Section (Section~\ref{sec:results}), by summing measurements of each run across the three datasets.

\subsection{Replication Package and Study Reproducibility}
\label{sec:rep_pkg}
To reproduce our experiment, we designed a comprehensive companion package (see Section~\ref{sec:intro}). To ensure rigorous replicability, extension, and scrutiny of our results, we tailor the replication package based on environments specific to each programming language. For Python, \textit{virtualenv} and \textit{conda} are used, while for MATLAB we bundle scripts and data in \textit{Live Scripts} and \textit{.mat} files. R findings reproducibility is supported by the \textit{packrat} and \textit{renv} environments, while for Java we rely on the \textit{Maven} dependency manager. Finally, C++ result replication is supported by rigorously documented build scripts and package dependencies.

\section{Results}
\label{sec:results}
In this section, we report the results of our experiment, namely the energy required to execute different AI algorithms implemented in various programming languages (see Section~\ref{sec:variables}). 

As documented in Section~\ref{sec:variables}, we focus separately on two distinct AI phases, namely the model training phase and the inference phase.

\subsection{Results $RQ_{1.1}$: Training Energy Efficiency of Programming Languages}
\label{sec:results_rq1}
\begin{figure*}[hbpt]
    \centering
    \includegraphics[width=2\columnwidth]{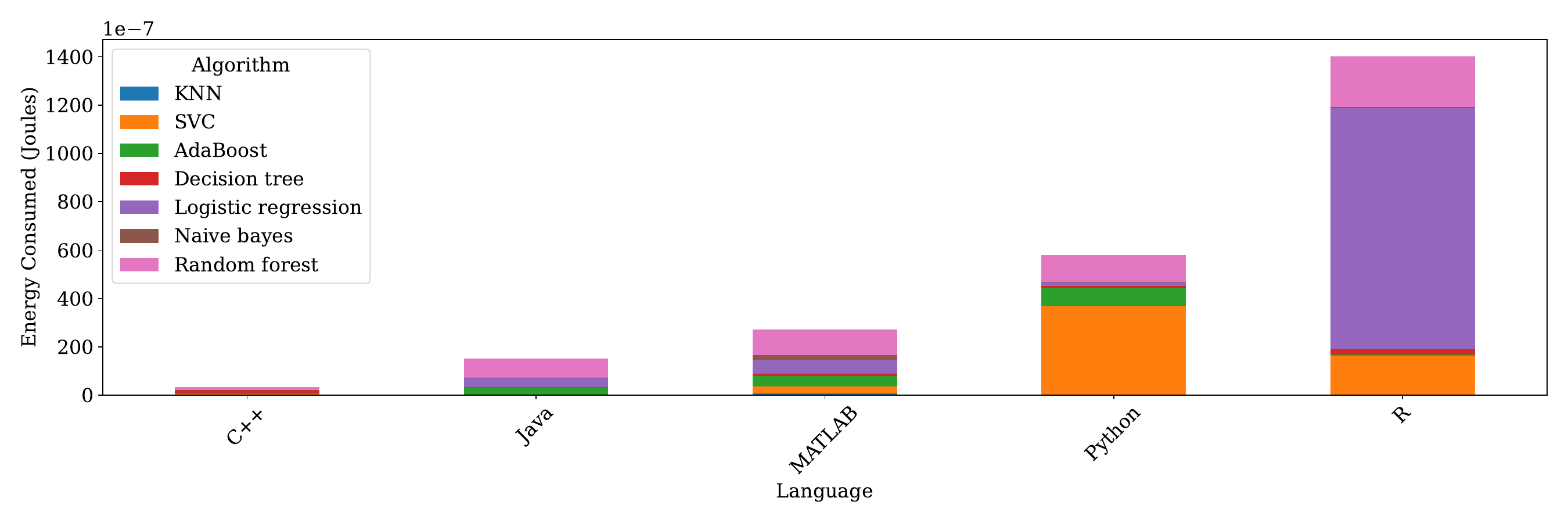}
    \caption{Energy Consumption by Language and Algorithm (Training Phase)}
    \label{fig:training-energy}
\end{figure*}

An overview of the cumulative energy consumed by the different programming languages during the training phase is depicted in Figure~\ref{fig:training-energy}.

As we can observe from Figure~\ref{fig:training-energy}, the overall energy consumed by the different programming languages during the training phase drastically differs. Overall, a clear ranking of energy consumption across the programming languages can be recognized, with C++ being the most energy efficient language, followed by Java, MATLAB, Python, and R. By considering the most and least energy efficient languages, namely C++ and R respectively, we note that in absolute terms R consumes 37 times more than C++ in the training phase. 

The energy efficiency difference between other languages is less remarked, with the minimum energy increment being between C++ and Java, the latter of which however still requires four times the energy consumed by C++.

Overall, regardless of the pair of languages considered, we deem the difference in energy consumed across different programming languages remarkable.

The described cumulative energy consumption figures provide fundamentally more insights when we zoom in into the energy consumed by the individual algorithms implemented in the different programming languages (see bar stacks of Figure~\ref{fig:training-energy}). Here we note that the implementation of a single algorithm can single-handedly lead a language to be drastically less energy efficient in absolute terms.

A prime example of this is the energy consumed by the logistic regression implementation in R, that contributes to a staggering 71.1\% of the total energy consumed by the language. Intuitively, one could assume that linear regression is \textit{per se} a very computational-intensive algorithm. However, by considering the other programming languages, we note that linear regression consistently ranks second or third in terms of training energy efficiency (out of seven algorithms), and in no other instance displays such a remarked difference in energy consumption compared to the other algorithms.

A similar case of profound impact of a single algorithm on total programming language energy consumption can be noted in the Python implementation of SVC, which contributes to 63.38\% of the total energy consumed by the language. Similarly, decision tree in Java accounts for approximately half (52.48\%) of the total energy consumed by such language.

A granular documentation of the energy consumed by each algorithm implemented in the different languages is provided in Table~\ref{tab:energy_consumption}. By considering the most energy efficient languages we note that, while C++ showcases the overall lowest cumulative energy consumption, Java is the language that provides the highest number of most energy efficient algorithm implementations (four out of seven algorithms, namely KNN, SVC, decision tree, and naive bayes, see Table~\ref{tab:energy_consumption}). Again, this fact is attributed to the major impact on total language energy consumption that single algorithm implementations can have.

On the other side of the training energy efficiency spectrum, while the high cumulative energy consumption of R is mostly due to its logistic regression implementation, R is also the programming language reporting the highest number of energy greedy algorithm implementations (three out of seven, namely decision tree, logistic regression, and random forest, see Table~\ref{tab:energy_consumption}). Despite the overall energy inefficiency of~R, two algorithms in this language, namely KNN and AdBoost, report the second less energy greedy implementation across all languages, with the latter algorithm consuming approximately six time less than its closest competitor (AdaBoost in Java).

\begin{table*}[hbpt]
\centering
\caption{Energy Consumption by Language and Algorithm in Joules (values multiplied by \(10^7\) for readability). Bold and underlined values represent respectively the lowest and highest value of a column.}
\begin{tabular}{lrrrrrrrr}
\toprule
\multicolumn{9}{c}{\textbf{Train energy consumption ($\mathbf{10^7}$ Joules)}} \\
\toprule
 & KNN & SVC & AdaBoost & Decision tree & Logistic regression & Naive bayes & Random forest & All\\ \midrule
C++ & 2.67 & 4.87 & \textbf{1.44} & 12.21 & \textbf{1.91} & 0.44 & \textbf{13.54} & \textbf{37.08} \\
Java & \textbf{0.34} & \textbf{0.38} & 32.76 & \textbf{0.33} & 38.31 & \textbf{0.35} & 80.06 & 152.54 \\
MATLAB & \underline{7.83} & 29.55 & 43.05 & 7.79 & 56.35 & \underline{21.12} & 106.46 & 272.14 \\
Python & 1.61 & \underline{366.85} & \underline{72.81} & 10.57 & 14.82 & 1.70 & 110.43 & 578.78 \\
R & 0.65 & 163.10 & 5.27 & \underline{21.74} & \underline{996.18} & 5.23 & \underline{209.05} & \underline{1401.23} \\
\toprule
\multicolumn{9}{c}{\textbf{Inference energy consumption ($\mathbf{10^7}$ Joules)}} \\
\toprule
 & KNN & SVC & AdaBoost & Decision tree & Logistic regression & Naive bayes & Random forest & All \\ \midrule
Java & \textbf{0.36} & \textbf{0.34} & 0.35 & 0.35 & 0.34 & \textbf{0.33} & \textbf{0.34} & \textbf{2.41} \\
C++ & 1.74 & 0.89 & 0\textbf{.27} & \textbf{0.21} & \textbf{0.17} & 0.69 & 0.78 & 4.75 \\
Python & \underline{13.70} & \underline{54.83} & 7.08 & 0.81 & 0.67 & 0.87 & 6.61 & 84.58 \\
R & 12.41 & 16.76 & \underline{24.94} & \underline{2.96} & 6.08 & 24.70 & 8.17 & 96.02 \\
MATLAB & 8.01 & 6.20 & 7.74 & 2.06 & \underline{6.67} & \underline{81.35} & \underline{17.63} & \underline{129.66} \\
\bottomrule
\end{tabular}
\label{tab:energy_consumption}
\end{table*}

\subsection{Results $RQ_{1.2}$: Inference Energy Efficiency of Programming Languages}
\label{sec:results_rq2}
\begin{figure*}[hbpt]
    \centering
    \includegraphics[width=2\columnwidth]{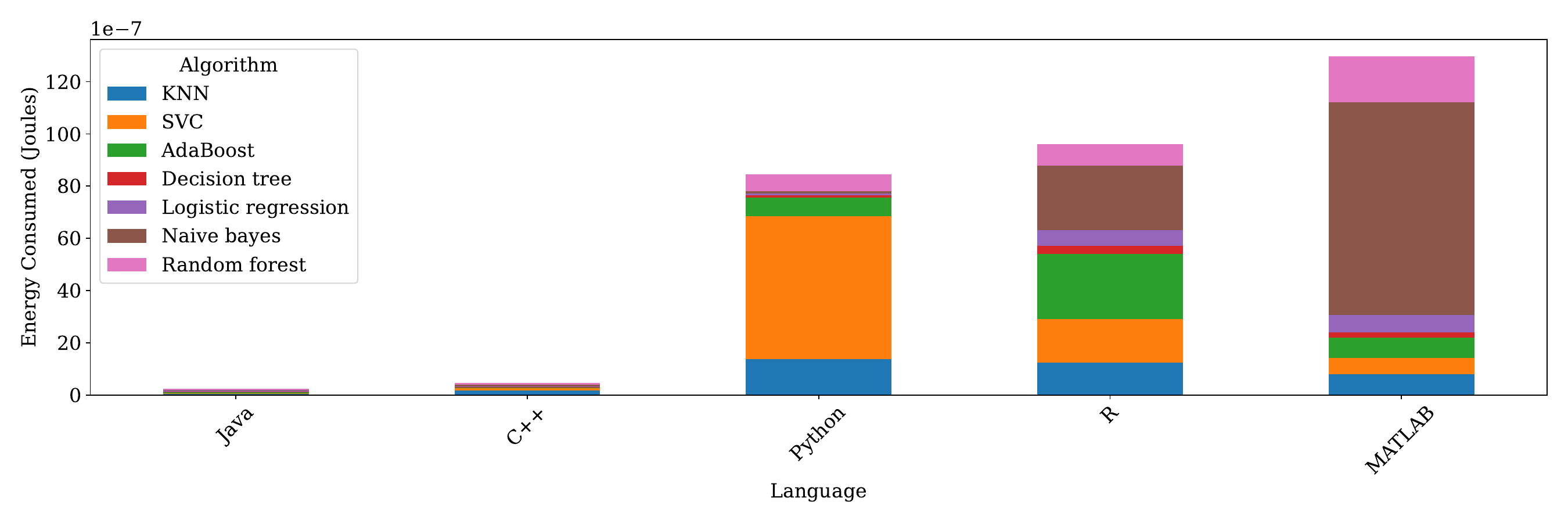}
    \caption{Energy Consumption by Language and Algorithm (Inference Phase)}
    \label{fig:inference-energy}
\end{figure*}

The energy consumed by the programming languages during the inference phase is depicted in Figure~\ref{fig:inference-energy}.

As we can observe in Figure~\ref{fig:inference-energy}, the energy efficiency of languages during inference drastically changes w.r.t. the training phase (see Figure~\ref{fig:training-energy}). In fact, not a single programming language  maintains its energy efficiency ranking across both phases. Further considerations on this matter, supported by joint discussion on the training and inference phases, are documented in our Discussion Section (Section~\ref{sec:discussion}).

When focusing on the inference phase, we note that Java results the programming language with the overall lowest cumulative energy consumption. Compared to the other languages considered in our experiment, Java consumes approximately two times less than the the second most energy efficient programming language (C++), and 54 times less than the most energy greedy language (R). 

A notable difference is present in the energy consumed by models implemented in Java and C++ when these are compared to the other languages. While the cumulative energy difference between Java and C++ is a 2x increase, by considering the third most energy efficient language (Python), such difference drastically increases by a 18x factor.

At a refined level of granularity, we can consider the energy consumed by each algorithm during inference (see bar stacks of Figure~\ref{fig:inference-energy} and Table~\ref{tab:energy_consumption}). As for the training phase, the implementation of single algorithms plays a crucial role in the overall energy efficiency of a language. As most notable example, SVC contributes to 64.8\% of the cumulative energy consumed by Python during the inference phase. Similarly, naive bayes in MATLAB contribute to 62.7\% of the total energy consumed by the MATLAB language. 

By considering the highest number of energy efficient algorithm implementations, we note that Java is not only the language consuming less cumulative energy, but also the language presenting the highest number of energy efficient algorithms (four out of seven, namely KNN, SVC, naive bayes, and random forest). C++ follows closely, with the rest of the most energy efficient algorithms implemented in such language (three out of seven, namely AdaBoost, decision tree, and logistic regression). 

The most energy greedy language in the inference phase, namely MATLAB, is also the one reporting the highest number of less energy optimized algorithms (three out of seven, namely logistic regression, naive bayes, and random forest). Python and R follow up closely, by presenting each two out of seven most energy greedy algorithm implementations.

As corollary note on the above reported results, also for the inference phase we are not able to identify algorithms that display an intrinsic energy-greediness, \ie specific algorithms that consume more when compared to the others regardless of the programming language they are implemented in.

\section{Discussion}
\label{sec:discussion}
In this section, we report our interpretation of the results presented in Section~\ref{sec:results}, with the aim of providing a clear and motivated answer to our RQs (see Section~\ref{sec:rqs}). 

Regarding the training phase, to explicitly answer $RQ_{1.1}$ (see Section~\ref{sec:rqs}), the impact of programming language on AI energy consumption is considerable (see Section~\ref{sec:results_rq1} for related results). From a minimum magnitude of four times an energy increment, up to a maximum of 37 times, choosing one programming language over another can have a profound influence on training energy efficiency. By considering the cumulative energy consumed by programming languages across algorithms, a clear ranking emerges: C++ is far more efficient than Java, which is more efficient than MATLAB, and following Python and R. Therefore, we conclude that, regardless of the pair of programming languages considered, each one is associate to a distinct cumulative energy consumption. 

\begin{highlight}
    \faLeaf~\textbf{Training Energy Efficiency of Programming Languages.} In terms of cumulative training energy consumption, the most efficient language is C++, followed by Java, MATLAB, Python, and R. C++ consumes for training 4 times less than Java, 7 times less than MATLAB, 15 times less than Python, and 37 times less than R.
\end{highlight}

Regarding the cumulative energy consumed in inference, a different ranking of programming language energy efficiency emerges, namely Java, C++, Python, R, and MATLAB. As observation on the cumulative energy consumption a noticeable gap in energy consumption between compiled / semi-compiled languages (C++, Java) and interpreted languages (Python, R, and MATLAB) is present. Overall by considering the inference phase, to explicitly answer $RQ_{1.2}$, we conclude that programming languages have a considerable impact on AI inference energy efficiency. 

\begin{highlight}
    \faLeaf~\textbf{Inference Energy Efficiency of Programming Languages.} By considering total inference energy consumption, the most efficient language is Java, followed by C++, Python, R, and MATLAB. Java consumes in inference 2~times less than C++, 35 times less than Python, 39 times less than R, and 54 times less than MATLAB.
\end{highlight}

By jointly considering training and inference, to explicitly answer our main $RQ$, we conclude that programming languages have a considerable impact on AI energy consumption. Regardless of the phase, C++ and Java are in vast majority of the cases the most energy efficient languages. Albeit Python, MATLAB, and R may use to various extents pre-compiled binary libraries, the overhead the interpreted languages \textit{per se} introduce cannot be completely evened out \textit{via} code optimization heuristics to improve their energy efficiency.

Our study takes a pragmatic stance on AI energy consumption of programming languages by considering popular AI libraries available ``in the wild'', rather than crafting \textit{ad-hoc} algorithm implementation as a purely academic exercise. In concrete terms, when choosing among different programming languages to optimize energy consumption, it is crucial to consider also which AI phase is most important in the specific context at hand. For example, MATLAB results to be rather efficient in training terms, while being considerably more energy-greedy during the inference phase. Therefore, as could expected, MATLAB is more suited for preliminary experimentation and proof of concept development, rather than for being deployed in a production environment at scale.

\begin{highlight}
    \faLeaf~\textbf{Green AI: Context matters.} Some languages are more energy efficient in training, others in inference. Which language to pick to optimize energy consumption highly depends on the specific application context considered.
\end{highlight}

Despite the reported clear-cut energy-efficiency rankings, which language to pick for AI tasks when software environmental sustainability is a primary concern is not trivial. In fact, when focusing at a refined level of granularity on different algorithms, which programming language is overall most energy efficient becomes fuzzy. Picking C++ or Java would with very high probability be a sound choice for energy efficiency, but such selection could lead to accidentally picking one of the most energy greedy algorithm implementations across all languages. For example, while C++ could be deemed overall as the most energy efficient language of the training phase, its decision tree implementation results to be the second most energy-greedy among all languages, making it a suboptimal language choice if only this specific algorithm is needed.

Perhaps as most important finding of this inquiry, we observe that the specific source code implementation of an algorithm plays a fundamental role in energy consumption, regardless of the programming language and algorithm considered. Therefore, while programming languages have a clear impact on AI energy consumption, anecdotally we conjecture that the particular source code implementation of an algorithm is the most determining factor for its energy consumption.

In a sense, this research serves as a cautionary tale, grounded in empirical data, warning to be wary about holistic statements such as ``Language X is the most energy efficient''. As observed in this study, while some general conclusions regarding programming language energy efficiency in AI can be drawn, we note that algorithm implementation could be the most determining factor of this results. Therefore we have to be wary not to wrongfully assume that one programming language is always more energy efficient than another, as this is empirically proven not to always be the case.

\begin{highlight}
    \faLeaf~\textbf{Algorithm implementation may be the most important factor of AI energy efficiency.} Algorithm implementation plays a fundamental role in AI energy consumption. Picking one language over another does not \textit{per se} guarantee the energy efficiency of a specific AI~algorithm.
\end{highlight}

As final consideration we want to mention the trade-off, neglected in this study, between energy efficiency and implementation ease. While programming languages lying at a lower level of abstraction might be more energy-efficient, implementing and maintaining AI algorithms in such languages might in some cases be or become unsustainable from a technical and/or economical point of view. The tradeoff between the environmental and the other dimensions of sustainability should be carefully evaluated, in order to avoid that an AI product becomes unmanageable just for the sake of energy efficiency. As last resort, if environmental sustainability is a primary target, the trade-off between energy efficiency and implementation ease should still be balanced by trying to optimize energy consumption without disrupting current development practices and technologies in place (\eg \textit{via} heuristics such as pre-compiled binary libraries implemented in more energy-efficient languages).

\begin{highlight}
    \faLeaf~\textbf{AI environmental sustainability needs to be balanced with implementation and maintenance ease.} Focusing solely on environmental sustainability is not an option. Trade-offs between environmental sustainability of AI and other sustainability dimensions, such as the technical and economic ones, should always be carefully considered. 
\end{highlight}

\section{Threats to Validity}
\label{sec:ttv}
In this section, we report the most prominent threats to validity that could have affected our results, by following the threats to validity categorization for controlled experiments outlined by Wholin~\etal~\cite{wohlin2012experimentation}, and the recent considerations on how to address threats to validity in software engineering research by Lago~\etal~\cite{lago2024threats}.

\subsection{Conclusion Validity}
In terms of internal validity, to mitigate potential threats caused by low statistical power, we repeat each experimental run 30 times (see Section~\ref{sec:data_collection}). Such mitigation strategy allows us also to strengthener our experimental design against threats to reliability of measures. As additional precaution against unreliable measurements, we use time series analysis to identify potential confounding factors polluting our experimental setting, \eg CPU throttling (see also Section~\ref{sec:data_collection}). 
To mitigate possible result fishing, in addition to our documented sampling strategy, we report our raw and processes results for independent scrutiny and interpretation (see Table~\ref{tab:energy_consumption} and the companion replication package linked in Section~\ref{sec:intro}). At the intersection of conclusion and external validity, the AI libraries chosen for this experiment (see Section~\ref{sec:procedure}) may constitute a threat to treatment implementation reliability. To mitigate this threat, we use for our experiment \textit{de facto} standard libraries for each programming language, such as \textit{scikit-learn}, \textit{mlpack}, and \textit{WEKA} (see Section~\ref{sec:procedure} and Section~\ref{sec:external_ttv}).

\subsection{Internal Validity}
Given the nature of our experiment, as no control group is used in our study, we focus on single group threats. Regarding potential history and maturation threats, confounding factors such as CPU throttling and hardware heat dissipation could  influence our measurements. As mitigation strategy (i) we consider for analysis the median values sampled across 30 subsequent execution, and (ii) conduct an \textit{a posteriori} data quality assurance process based on time series analysis of the measurements (see Section~\ref{sec:data_collection}). Regarding instrumentation threats, as we rely on a library providing native kernel-level energy consumption measurements (see Section~\ref{sec:procedure}), we deem our energy measurement to be both accurate and reliable.

\subsection{Construct Validity}
A major threat to construct validity lies in the inadequate preoperational explication of our construct. To mitigate this threat, we actively refrain from adopting convoluted derived metrics of software environmental sustainability such as carbon footprint or software carbon intensity\footnote{\url{https://github.com/Green-Software-Foundation/sci}. Accessed 1st November 2024.}. As further detailed in Section~\ref{sec:variables}, sustainability metrics of such nature would make our results very hard to be compared, reproduced, and generalized. For this reason, we use as construct the most simple and closest to bare metal interpretation of software environmental sustainability we are aware of, namely hardware energy consumption. 

As \textit{a posteriori} discovered construct validity threat influencing our study, the construct level influences our results, as rather than focusing solely on programming languages, a more correct scoping to achieve our research goal (see Section~\ref{sec:rqs}) would be to understand which algorithm implementation choices in different programming languages are more energy efficient. To partially address this threat, we thoroughly discuss the impact of algorithm implementation on energy consumption, highlighting such result also as our most relevant finding while leaving a complete mitigation of this threat as future work.

\subsection{External Validity}
\label{sec:external_ttv}
In terms of external validity, the results presented in this study should not be deemed in any way as conclusive or generalizable in absolute terms. 

Albeit as mitigation strategy the experimental objects are chosen by following selection criteria that enforce heterogeneity, the datasets represent only a fraction of the application scenarios the chosen algorithms can be used for. Future research should be conducted to study if, and in case to what extent, the results presented in this study hold when other datasets are considered.

Similarly, the empirical experiment conducted for this study focuses on seven  algorithms representative of different algorithmic families (\eg SVC for support vector machines). The choice to focus on a set of representative algorithms constituted a trade-off of external validity in favour of internal validity, which allowed us to collect sound measures \textit{via} a high number of reruns, while keeping the experiment feasible in terms of execution time. To strengthener the external validity of this study, future research should consider further algorithms of each algorithmic family.

As prominent related threat, in our experiment a single implementation is considered for each pair of algorithm and programming language (see Section~\ref{sec:variables}). To mitigate this threat, we select the out of the box implementation of the algorithms provided by \textit{de facto} standard libraries of each programming language. As for the previous threat, this constituted a trade-off between internal and external validity. To strengthen the external validity of the presented results, future research should consider different implementations of an algorithm in the same programming language.

\section{Related Work}
In this section, we discuss the researches related the closest to this study. To the best of our knowledge the topic we consider, namely the energy efficiency of different programming languages in the context of AI, constitutes a research gap in the existing Green AI body of literature~\cite{verdecchia2023systematic}. 

The most relevant related work is potentially the work of Georgiou~\etal~\cite{georgiou2022green}, which explores the energy efficiency of two Python deep learning frameworks. In contrast to such study, we focus on traditional AI rather than deep learning, and consider different programming languages, rather than different frameworks of one programming~language. 

As other closely related literature to this study, a set of researches investigate the impact that programming languages have on software energy consumption, albeit without a focus on AI. Pereira~\etal~\cite{pereira2017energy, pereira2021ranking} explore how energy, time, and memory relate across different programming languages by considering different computing scenarios (\eg DNA-matching, binary tree operations, and packing problems). Georgiou~\etal~\cite{10.1145/3139367.3139418} instead study how different compiled, semi-compiled, and interpreted languages compare in terms of energy-consumption on a set of classic programming problems taken from the Rosetta Code chrestomathy repository\footnote{\url{https://rosettacode.org}. Accessed 1st November 2024.}. 
In a more recent work by Gordillo~\etal~\cite{Gordillo2024}, the energy efficiency of different programming languages is evaluated on five classic programming problems, such as binary tree operations and n-body simulations. By considering more fine-grained contexts, Chandra~\etal~\cite{chandra2019impact} analyze programming language energy efficiency by focusing on different sorting algorithms,  Mahadevappa and Figueira~\cite{9612479} on mobile applications, and Maleki~\etal~\cite{maleki2017understanding} on design patterns. As main difference to all of the above mentioned studies, the focus of this research is investigating the impact that different programming languages can have in the context of AI.

% by considering the libraries we deem most popular for each language. This process is intrinsically threatened by subjective biases. To strengthen both internal and external validity, we plan and encourage other researchers to experiment with additional AI libraries providing different algorithm implementations.

\section{Conclusions and Future Work}
\label{sec:conclusion}
In this study, we report a controlled empirical experiment conducted to understand the impact of different programming languages on the energy consumption of AI. The research considers five programming languages (C++, Java, Python, MATLAB, and R), seven AI algorithms (KNN, SVC, AdaBoost, decision tree, logistic regression, naive bayses, and random forest), and two AI phases (training and inference). 

From the results collected in this study, we draw the following conclusion. First,  compiled and semi-compiled languages (C++, Java) are overall more energy efficient than interpreted languages (Python, R, and MATLAB), with an overall energy consumption going up to a maximum of a 54 times increase when using interpreted languages. Second, energy efficiency of programming languages can vary considerably when we consider the training or the inference phase. Third, despite potentially using pre-compiled libraries, the overhead introduced by interpreted languages might not be completely evened out \textit{via} code optimizations. Finally, algorithm implementation might be the most determining factor when considering the energy efficiency of AI algorithms.

As a word of warning in interpreting the collected results, and as underlying message these entail, the findings do not necessarily suggest to refactor codebases into more energy efficient languages. Rather, the results hint to the importance that algorithm implementation holds in energy efficiency, regardless of the language considered. In addition, tradeoffs between the environmental sustainability dimensions and other ones, such as technical and economic sustainability, should be carefully considered when optimizing AI energy consumption. 

As future work, we are eager to mitigate the threats to validity of our study (see Section~\ref{sec:ttv}), firstly by considering a wider range of experimental objects and languages. In addition, we are also curious to measure how energy consumption compares across libraries implemented in the same programming language. Finally, we are also interested to conduct studies based on the same research method in other AI contexts, with the ultimate goal of making, through systematic research steps, AI more environmentally sustainable.

\section*{Carbon Footprint of This Study}
The carbon footprint required to run the experiment conducted for this study is approximately 42.9 grams of $CO_2$, the same amount required to power an average electric car for 0.39 kilometers (0.24 miles)~\cite{zhao2023quantifying}.

\bibliographystyle{ieeetr}
\bibliography{bibliography}

\end{document}